\documentclass[aps,showpacs,showkeys,10pt]{revtex4}
\usepackage{graphicx}
\usepackage{epsfig}
\usepackage{bm}
\usepackage{amsmath}
\usepackage{amssymb}
\usepackage{graphics}
\allowdisplaybreaks

\begin{document}

\title{A New Fracture Function Approach to QCD Initial State Radiation}

\author{Federico A. Ceccopieri}
\email{ceccopieri@fis.unipr.it}
\affiliation{Dipartimento di Fisica, Universit\`a di Parma,\\
Viale delle Scienze, Campus Sud, 43100 Parma, Italy}
\author{Luca Trentadue}
\email{luca.trentadue@cern.ch}
\affiliation{Dipartimento di Fisica, Universit\'a di Parma,
INFN Gruppo Collegato di Parma, Viale delle Scienze, Campus Sud, 
43100 Parma, Italy}
\begin{abstract}
Ordinary fracture functions, describing hadrons production 
in the deep inelastic scattering target fragmentation region,
are generalized to account for the production of hadrons in arbitrary number,
thus offering a renewed framework for dealing with QCD initial state radiation.
We also propose a new jet-like observable 
which measures beam remnants and low-$p_{\perp}$ scattering fragments 
and  derive its QCD evolution equations by using Jet Calculus.
Possible implications for semi-inclusive deep inelastic scattering and hadron-hadron reactions 
are shortly discussed.  
\end{abstract}

\maketitle

\section{Introduction}
\label{sec:intro}

The fracture functions approach~\cite{Trentadue_Veneziano} allows, 
within the framework of perturbative Quantum-Chromodynamics, 
to describe low-$p_{\perp}$ target fragmentation 
along with semi-hard radiation produced by  the active parton 
partecipating the hard, hadron-initiated, scattering.
Fracture functions are represented as non-perturbative distributions 
that, once properly generalized to describe 
multi-particles production, allow an improved description of QCD 
initial state radiation.
In order to specify the connection of those distributions with the standard 
perturbative QCD approach to the dynamics of initial state, let us consider a fast incoming 
hadron entering into a high energy reaction.  
In a QCD-improved parton model approach, the hadron can be represented as 
a cloud of quasi-real partons which may fluctuate into far off-shell states.
If no hard scattering takes place, the virtuality of every single 
space-like parton would lead the complete cascade to reassamble.  
If instead a hard scattering occurs, the virtual probe, being able to resolve 
parton fluctuations in the hadron up to scales of the order $\mathcal{O}(1/Q^2)$,
knocks out a parton from the incoming hadron. 
As a result, the remaining off-shell partons in the initial state cascade
cannot reassamble themselves any longer and therefore materialize 
by starting a partonic initial state shower.  
This radiation piles up in the final state with the one generated by the struck parton time-like decay 
in the region of $\mathcal{O}(Q^2)$ transverse momenta.
While the latter, due to its detailed knowledge from $e^+ e^-$ process, 
is under better theoretical control, the former is at present 
still under study, especially in the kinematical 
regime of present and forthcoming hadron colliders. 

In a standard QCD approach to hard processes, initial state radiation is inclusively summed over
by using DGLAP evolution equations~\cite{DGLAP}. 
The complementary approach in which initial state radiation is unintegrated, 
or only partially integrated, gives therefore access to a more detailed inspection of
space-like parton dynamics at high energy. 
The aim of this work is thus
to introduce multi-particle semi-inclusive fracture functions,  
which can describe the unintegrated initial state radiation, and 
to give the corresponding QCD evolution equations.   

We use the general theoretical framework of  
Jet Calculus, originally proposed in Ref. \cite{KUV} in order to describe jet 
fragmentation. Within this framework, a highly virtual time-like parton, 
generated in a hard process, 
degrades its virtuality emitting a tree-like cascade of coloured quanta of 
lower mass until the non-perturbative limit is reached and hadronization takes place.
Within a leading logarithmic approximation, n-particle cross-sections are given, 
in the portion of phase space specified by the jet and neglecting for simplicity
correlated fragmentations, in terms of single-hadron fragmentation functions.  
The tree-like structure of the cascade is dictated by leading logarithmic enhancements 
to the cross-sections given by planar diagrams in the collinear limit.
Such enhancements are resummed at all order by DGLAP-like evolution equations \cite{DGLAP}.
\newpage
In the following we apply the same techniques in the space-like case as appropriate 
for the description of initial state radiation. 
In order to simplify the notation it is convenient to replace the evolution variable $Q^2$ with
\begin{equation}
\label{Y}
Y=\frac{1}{2\pi \beta_0}\ln \Big[ \frac{\alpha_s(\mu_R^2)}{\alpha_s(Q^2)}\Big],
\;\;\;\; dY = \frac{\alpha_s(Q^2)}{2\pi}\frac{dQ^2}{Q^2}\,.
\end{equation}
At LL accuracy the running of $\alpha_s$ is taken into account at one loop by using 
\begin{equation}
\alpha_s(Q^2)=\frac{1}{\beta_0 \ln (Q^2/\Lambda_{QCD}^2)},
\end{equation}
where $\mu_R^2$ is the renormalization scale, $\Lambda_{QCD}^2$ 
is the QCD infrared scale, $\beta_0=(11 C_A -2n_f)/12\pi$ 
is the one loop $\beta$-function parameter and $C_A$ and $n_f$ are the number of colours and 
flavours respectively. We then define the variable $y$
\begin{equation}
\label{y}
y=\frac{1}{2\pi \beta_0}\ln \Big[ \frac{\alpha_s(Q_0^2)}{\alpha_s(Q^2)}\Big],
\end{equation}
where $Q_0^2$ and $Q^2$ stand for two arbitrary perturbative scales. We next 
introduce the perturbative parton-to-parton evolution function $E_i^j(x,y)$ which
expresses the probability of finding a parton 
$j$ at the scale $Q^2$ with a momentum fraction $x$ of the 
parent parton $i$ at the scale $Q_0^2$. 
The function $E_i^j(x,y)$ satisfies a DGLAP-type evolution equation \cite{KUV}
\begin{equation}
\label{E_evo}
\frac{\partial}{\partial y}E_i^j(x,y)=\int_x^1 \frac{du}{u} P^{j}_{k}(u) E_i^k
\Big(\frac{x}{u},y \Big)\,,
\end{equation} 
where $P^{j}_{k}(u)$ are the space-like splitting functions \cite{DGLAP} and 
a sum over the parton index $k$ is understood.
The daughter partons radiated in the evolution process of the active parton,
\textsl{i.e.} the virtual cascade of Sec.~\ref{sec:intro}, 
are inclusively summed by eq.~(\ref{E_evo}). The evolution equations can be iteratively 
solved by using the initial condition 
\begin{equation}
\label{ic}
E_i^j(x,y)|_{y=0}=\delta_{i}^{j}\delta(1-x) \,. 
\end{equation}
The function $E_i^j(x,y)$ resums large collinear logarithms of the type 
$\alpha_s^n \ln^n (Q^2/Q_0^2)$. Neglecting for simplicity the running of $\alpha_s$,
its expansion at first order in fact reads 
\begin{equation}
\label{E_LL}
E_i^j(x,y)\equiv E_i^j(x,Q_0^2,Q^2)\simeq\delta_{i}^{j}\delta(1-x)+
\frac{\alpha_s}{2\pi}P_{i}^{j}(x)\ln \frac{Q^2}{Q_0^2}\,+\mathcal{O}(\alpha_s^2).
\end{equation}
The $E$'s function in LLA therefore well describe the emission of partons that are soft
or close to the directions of the emitting parton, while large angle emissions
should be included via exact matrix elements. Furthermore,
the $E$'s function satisfies the following \textit{renormalization group} property:
\begin{equation}
\label{group_prop}
E_i^j(x,Q_0^2,Q^2)=\int_x^1\frac{dw}{w}E_i^k(x/w,Q_0^2,Q_i^2)E_k^j(w,Q_i^2,Q^2)\,,
\end{equation}
where the scale $Q_i^2$ is such that $Q_0^2<Q_i^2<Q^2$. This property is easily
verified once an $\alpha_s$-expansion is performed on both sides of eq.(\ref{group_prop}).


\section{Evolution equations for ordinary fracture functions}
\label{sec:M1}
Ordinary fracture functions have been introduced in Ref.~\cite{Trentadue_Veneziano} in order
to give a QCD-based description of semi-inclusive Deep Inelastic Scattering in the target 
fragmentation region. $M^i_{h/P}(x,z,Q^2)$ represents the conditional probability 
of finding at a given scale $Q^2$ a parton $i$ with momentum fraction $x$  
of the incoming hadron momentum $P$ while a hadron $h$ with momentum fraction $z$ is detected. 
All-order factorizazion of collinear and soft singularities 
into $M^i_{h/P}(x,z,Q^2)$ were demonstrated in 
Refs.~\cite{Fact_M_coll,Fact_M_soft}. In Ref.~\cite{Graudenz}, 
a fixed order $\mathcal{O}(\alpha_s)$ calculation 
showed explicitely that the additional collinear singularities occuring  
when partons are produced in the remnant direction can be properly renormalized only introducing 
fracture functions. In view of the generalization performed in the next Section, 
we briefly recall the derivation of the $M^i_{h/P}(x,z,Q^2)$ evolution equation.
Let us consider the DIS one-particle inclusive cross-sections $l+P\rightarrow l' + h + X $
in the target framentation region, $t=-(P-h)^2\ll Q^2$,
\begin{equation}
\label{sigma_T1}
\sigma_T=\int  \frac{du}{u} M^i_{h/P}(u,z,Q^2)\, \hat{\sigma}_i(x/u,Q^2)\,.
\end{equation}
The cross-sections is expressed as a convolution of the fracture functions 
with the point-like partonic cross-sections $\hat{\sigma}^i$. 
Fixed order calculations shows singularities, 
as already said, when the emitted parton is collinear to the hadron remnant. 
The structure of singularities is \textsl{rephrased} in a Jet Calculus approach by summing
over all combinations of distributions which can give the desidered final state. 
Therefore we may write 
\begin{eqnarray}
\label{M1_def}
M^{j}_{h/P}(x,z,Y)&=&M_{A,h/P}^{j}(x,z,Y)+M_{B,h/P}^{j}(x,z,Y)\,,\\
M_{A,h/P}^{j}(x,z,Y)&=&\int_{x}^{1-z}\frac{dw}{w}E_i^j\Big(\frac{x}{w},Y-y_0\Big)
M_{A,h/P}^{i}(w,z,y_0)\,,\\
M_{B,h/P}^{j}(x,z,Y)&=&\int_{y_0}^{Y}dy\int_{x+z}^{1}\frac{dw}{w^2}
\int_{\frac{x}{w}}^{1-\frac{z}{w}}\frac{du}{u(1-u)}\cdot\nonumber\\
&&\cdot E_k^j\Big( \frac{x}{wu},Y-y \Big) \hat{P}_{i}^{kl}(u)
D_l^h\Big(\frac{z}{w(1-u)},y\Big)F_P^i(w,y)\,.
\end{eqnarray}
\begin{figure}
\label{M1fig}
\begin{center}
\epsfig{file=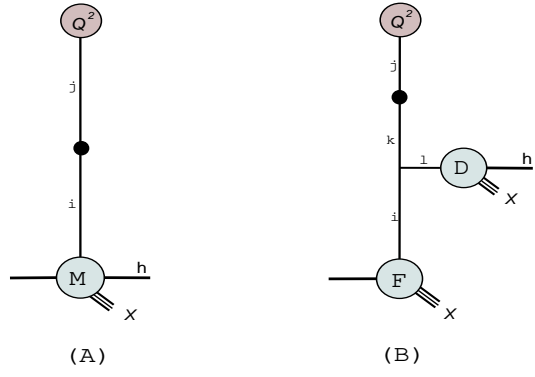,width=7cm,height=5cm,angle=0}
\caption{\small{The $A$ and $B$ term of eq.(\ref{M1_def}). 
The black blob denotes the parton-to-parton evolution function $E$.
Partons indices are indicated and at each triple-line 
vertex is associated a real AP splitting functions $\hat{P}(u)$. The diagrams 
are at the amplitude square level. The top parton line enters the hard $(Q^2)$ scattering
indicated by the bright blob.}}
\end{center}
\end{figure}
Both terms are depicted in Fig.~1.
The variable $y_0$ corresponds to an arbitrary factorization scale and $F_P^i$ and $D_l^h$ stands  
respectively for the structure function of a parton $i$ inside a proton, and 
the fragmentation function of a parton $l$ into the hadron $h$.  
$\hat{P}_{i}^{kl}(u)$ are the real AP splitting functions \cite{KUV}.
The $A$-term stems for direct hadron production from a fracture function, while the $B$-term, 
sometimes called perturbative, takes into account the production of $h$ 
by a timelike cascade of the daughter parton $l$. The scale associated to this branching, 
being not observable, has been integrated over.
The integration limits in eq.~(\ref{M1_def})
are fixed by requiring momentum conservation. 
The evolution equations are easily obtained by deriving  eq.(\ref{M1_def}) with respect to to $Y$. 
One has
\begin{eqnarray}
\label{MAB}
\frac{\partial}{\partial Y}M_{A,h/P}^{j}(x,z,Y)&=&
\int_{\frac{x}{1-z}}^{1}\frac{du}{u}P_i^j(u)M_{A,h/P}^{j}(x/u,z,Y)\,,\\
\frac{\partial}{\partial Y}M_{B,h/P}^{j}(x,z,Y)&=&
\int_{\frac{x}{1-z}}^{1}\frac{du}{u}P_i^j(u)M_{B,h/P}^{j}(x/u,z,Y)+\\
&&\int_{x}^{\frac{x}{x+z}}\frac{du}{u}\frac{u}{x(1-u)}
\hat{P}_{i}^{jl}(u) D_l^h\Big(\frac{zu}{x(1-u)},Y\Big)F_P^i(x/u,Y)\,.
\end{eqnarray}
Restoring the familiar variable $Q^2$, we obtain the evolution equation for 
$M_{h/P}^{j}(x,z,Q^2)$:
\begin{eqnarray}
\label{evoM1}
Q^2\frac{\partial}{\partial Q^2}M_{h/P}^{j}(x,z,Q^2)&=&\frac{\alpha_s(Q^2)}{2\pi}
\int_{\frac{x}{1-z}}^{1}\frac{du}{u}P_i^j(u)M_{h/P}^{i}(x/u,z,Q^2)\,+\\
&&\frac{\alpha_s(Q^2)}{2\pi}\int_{x}^{\frac{x}{x+z}}\frac{du}{u}\frac{u}{x(1-u)}
\hat{P}_{i}^{jl}(u) D_l^h\Big(\frac{zu}{x(1-u)},Q^2\Big)F_P^i(x/u,Q^2)\,.\nonumber
\end{eqnarray}
\newpage
$M_{h/P}^{j}(x,z,Q^2)$ does not depend on the  factorization scale 
and satisfies its own s-channel sum rule \cite{Trentadue_Veneziano} :
\begin{eqnarray}
\label{M_sum_rule}
\sum_{h}\int dz\, z\, M^j_{h/P}(x,z,Q^2)&=&(1-x)F^j_P(x,Q^2)\,.
\end{eqnarray}
At the phenomenological level, fracture functions have 
been shown in Ref.~\cite{Mpheno} to well reproduce at the same time 
both HERA diffractive and leading proton data, thus convalidating a common perturbative
QCD approach to these particular classes of semi-inclusive processes. 
 

\section{Evolution equations for di-hadrons fracture functions}
\label{sec:M2}

Let us consider a double-inclusive Deep Inelastic Scattering process
$l+P\rightarrow l'+h_1 + h_2 +X$ where two detected hadrons,
$h_1$ and $h_2$, have both $t_{i=1,2}=-(P-h_i)^2\ll Q^2$.
In analogy with eq.~(\ref{sigma_T1}) we may write the corresponding 
double-inclusive cross sections as 
\begin{equation}
\label{sigma_T2}
\sigma_T=\int  \frac{du}{u} M^j_{h_1,h_2/P}(u,z_1,z_2,Q^2)\, \hat{\sigma}_j(x/u,Q^2)\,.
\end{equation}
$M_{h_1,h_2/P}^i(x,z_1,z_2,Q^2)$ 
gives the conditional probability of finding an active quark $i$ with momentum fraction $x$
of the incoming hadron momentum P while two secondary hadrons with fractional energy $z_1$ and $z_2$
are detected. 
Evolution equations for di-hadron fracture functions $M_{h_1,h_2/P}^i(x,z_1,z_2,Q^2)$
can be obtained generalizing the derivation outlined in the previous Section.
$M_{h_1,h_2/P}^i(x,z_1,z_2,Q^2)$ is therefore given by the incoherent sum of all combination 
of distributions which can give $h_1$ and $h_2$ in the final state, see Fig.~2:
\begin{equation}
\label{M2sum}
M_{h_1,h_2/P}^j(x,z_1,z_2,Y)=\sum_{X=A,B,C,D}M_{X,h_1,h_2/P}^j(x,z_1,z_2,Y)\,.
\end{equation}
Explicitely the four contributions read
\begin{eqnarray}
\label{M2ABCD}
M_{A,h_1,h_2/P}^j(x,z_1,z_2,Y)&=&\int_{x}^{1-z_1-z_2}\frac{dw}{w}E^j_i
\Big(\frac{x}{w},Y-y_0\Big)M_{A,h_1,h_2/P}^i(w,z_1,z_2,y_0)\,,\\
M_{B,h_1,h_2/P}^j(x,z_1,z_2,Y)&=&\int_{y_0}^{Y}dy \int_{x+z_1}^{1-z_2}\frac{dw}{w^2}\int_{x/w}^{1-z_1/w}
\frac{du}{u(1-u)} E_k^j\Big(\frac{x}{wu},Y-y\Big)\hat{P}_i^{kl}(u)\cdot\nonumber\\
&& M_{A,h_2/P}^i(w,z_2,y)D_l^{h_1}\Big( \frac{z_1}{w(1-u)},y\Big)+(h_1,z_1)\leftrightarrow(h_2,z_2)\,,\\
M_{C,h_1,h_2/P}^j(x,z_1,z_2,Y)&=&\int_{y_0}^{Y}dy_2 \int_{y_2}^{Y}dy_1
\int_{x+z_1+z_2}^{1}\frac{dw}{w^2}\int_{x+z_1}^{w-z_2}\frac{dx_2}{x_2^2}
\int_{x/x_2}^{1-z_1/x_2}\frac{du_1}{u_1(1-u_1)}\hat{P}_m^{n l_1}(u_1)\cdot \nonumber\\
&&\int_{x_2/w}^{1-z_2/w}\frac{du_2}{u_2(1-u_2)}\hat{P}_i^{k l_2}(u_2)
E_k^m\Big(\frac{x_2}{wu_2},y_1-y_2\Big)E_n^j\Big(\frac{x}{u_1 x_2},Y-y_1\Big)\cdot\nonumber\\
&&D_{l_2}^{h_2}\Big( \frac{z_2}{w(1-u_2)},y_2\Big)
D_{l_1}^{h_1}\Big( \frac{z_1}{x_2 (1-u_1)},y_1\Big)F_P^i(w,y_1)+(h_1,z_1)\leftrightarrow(h_2,z_2)\,,\\
M_{D,h_1,h_2/P}^j(x,z_1,z_2,Y)&=&\int_{y_0}^{Y}dy 
\int_{x+z_1+z_2}^{1}\frac{dw}{w}\int_{x/w}^{1-(z_1+z_2)/w}
\frac{1}{w^2(1-u)^2}
\frac{du}{u} E_k^j\Big(\frac{x}{wu},Y-y\Big)\cdot\nonumber\\
&&\hat{P}_i^{kl}(u)F_P^i(w,y)D_l^{h_1,h_2}\Big( \frac{z_1}{w(1-u)}, \frac{z_2}{w(1-u)},y\Big)\,.
\end{eqnarray} 
\begin{figure}
\label{M2fig}
\begin{center}
\epsfig{file=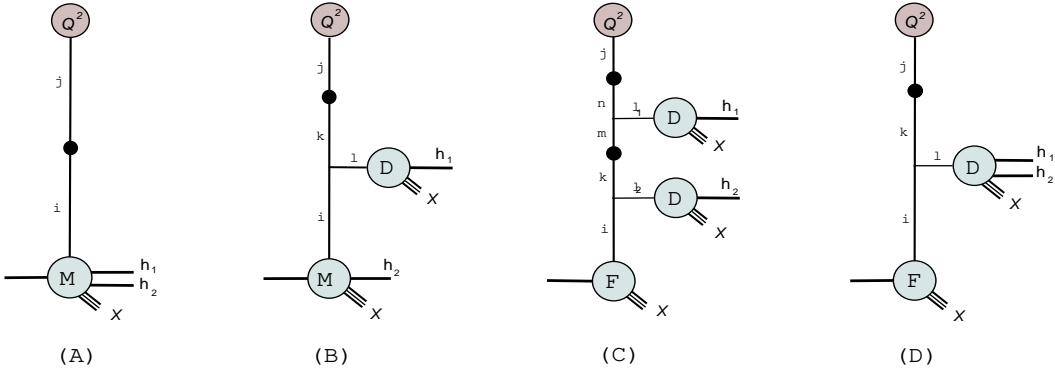,width=14cm,height=5cm,angle=0}
\caption{\small{The four terms of eq.~(\ref{M2sum}). 
The black blob denotes the parton-to-parton evolution function $E$.
Partons indices are shown and at each triple-line 
vertex is associated a real AP splitting functions $\hat{P}(u)$. The diagrams 
are at the amplitude square level. The top parton line enters the hard $(Q^2)$ scattering
indicated by the bright blob.}}
\end{center}
\end{figure}
The $A$-term is the direct convolution of a di-hadron 
fracture functions with the function $E$. 
The $B$-term involves one-hadron fracture functions and fragmentation functions.  
The third and more involved $C$-term is of a ladder type and contain two $y-$integration 
because the two hadrons are emitted by daughters of the active parton ordered by 
virtualities corresponding to $y_1$ and $y_2$. 
The $D$-term involves di-hadron fragmentation functions which
obeys its own inhomogeneous-type evolution equations \cite{KUV,Deflorian_Vanni}. 
The $B$ and $C$ terms also contain twin diagrams with $h_1$ and $h_2$ 
interchanged, as indicated by $(h_1,z_1)\leftrightarrow (h_2,z_2)$. 
All the integration limits in the convolution integrals are fixed by requiring momentum conservation.
Taking the $Y$-derivative of all terms and making repeated use of 
the evolution equation, eq.~(\ref{E_evo}), and of its intial condition, eq.~(\ref{ic}), 
we obtain
\newpage
\begin{eqnarray}
\label{M2A}
\frac{\partial}{\partial Y} M_{A,h_1,h_2/P}^j
&=&\int_{\frac{x}{1-z_1-z_2}}^{1}\frac{du}{u}P_i^j(u)M_{A,h_1,h_2/P}^i(x/u,z_1,z_2,Y)+\,,\\
\label{M2B}
\frac{\partial}{\partial Y} M_{B,h_1,h_2/P}^j
&=&\int_{\frac{x}{1-z_2}}^{\frac{x}{x+z_1}}
\frac{du}{u}\frac{u}{x(1-u)}\hat{P}^{jl}_i(u)M_{A,h_2/P}^i(x/u,z_2,Y)D_l^{h_1}
\Big( \frac{z_1 u}{x(1-u)},Y\Big)+ \nonumber\\
&&+(h_1,z_1)\leftrightarrow(h_2,z_2)+
\int_{\frac{x}{1-z_1-z_2}}^1\frac{du}{u}P_i^j(u)M_{B,h_1,h_2/P}^i(x/u,z_1,z_2,Y)\,,\\
\label{M2C}
\frac{\partial}{\partial Y} M_{C,h_1,h_2/P}^j
&=&\int_{\frac{x}{1-z_2}}^{\frac{x}{x+z_1}}
\frac{du}{u}\frac{u}{x(1-u)}\hat{P}_m^{j l_1}(u)M_{B,h_2/P}^m(x/u,z_2,Y)D_{l_1}^{h_1}
\Big( \frac{z_1 u}{x(1-u)},Y\Big)\nonumber\\
&&+(h_1,z_1)\leftrightarrow(h_2,z_2)
+\int_{\frac{x}{1-z_1-z_2}}^1\frac{du}{u}P_i^j(u)M_{C,h_1,h_2/P}^i(x/u,z_1,z_2,Y)\,,\\
\label{M2D}
\frac{\partial}{\partial Y} M_{D,h_1,h_2/P}^j
&=&\int_{x}^{\frac{x}{x+z_1+z_2}}
\frac{du}{u}\frac{u^2}{x^2(1-u)^2}\hat{P}_i^{jl}(u)F^i_P(x/u,Y) 
D_l^{h_1,h_2}\Big( \frac{z_1 u}{x(1-u)}, \frac{z_2 u}{x(1-u)},Y\Big)\nonumber\\
&&+\int_{\frac{x}{1-z_1-z_2}}^1\frac{du}{u}P_i^j(u)M_{D,h_1,h_2/P}^i(x/u,z_1,z_2,Y).
\end{eqnarray}
Terms containing ordinary splitting functions $P_i^j(u)$ sum up
to give a homogeneous term for the evolution equation for $M_{h_1,h_2/P}$. 
Terms $M_{A,h_2/P}$ and $M_{B,h_2/P}$ in first line of eq.~(\ref{M2B}) and eq.~(\ref{M2C}) 
combine to give $M_{h_2/P}$, eq.~(\ref{M1_def}).
We can thus finally write the evolution equation for $M_{h_1,h_2/P}$ as:
\begin{eqnarray}
&&Q^2\frac{\partial}{\partial Q^2} M_{h_1,h_2/P}^j(x,z_1,z_2,Q^2)=
\frac{\alpha_s(Q^2)}{2\pi}\int_{\frac{x}{1-z_1-z_2}}^{1}
\frac{du}{u}P_i^j(u)M_{h_1,h_2/P}^i(x/u,z_1,z_2,Q^2)+\nonumber\\
&&+\frac{\alpha_s(Q^2)}{2\pi}\int_{\frac{x}{1-z_2}}^{\frac{x}{x+z_1}}
\frac{du}{u}\frac{u}{x(1-u)}\hat{P}_i^{jl}(u)M_{h_2/P}^i(x/u,z_2,Q^2)
D_l^{h_1}\Big( \frac{z_1 u}{x(1-u)},Q^2\Big)+(h_1,z_1)\leftrightarrow(h_2,z_2)\,+\\
&&+\frac{\alpha_s(Q^2)}{2\pi}\int_{x}^{\frac{x}{x+z_1+z_2}}
\frac{du}{u}\frac{u^2}{x^2(1-u)^2}\hat{P}_i^{jl}(u)F^i_P(x/u,Q^2) 
D_l^{h_1,h_2}\Big( \frac{z_1 u}{x(1-u)}, \frac{z_2 u}{x(1-u)},Q^2\Big)\,.\nonumber
\end{eqnarray}
As in the case of ordinary fracture functions $M_{h/P}$, eq. (\ref{M_sum_rule}), 
also di-hadron fracture functions $M_{h_1,h_2/P}$ can be shown to be independent 
on the factorization scale variable $y_0$ and satisfie s-channel sum rules:
\begin{eqnarray}
&&\sum_{h_2}\int \,dz_2 \, z_2 \, M_{h_1,h_2 /P}^{i}(x,z_1,z_2,Q^2)
=(1-x-z_1)M_{h_1/P}^i(x,z_1,Q^2)\,,\\
&&\sum_{h_1,h_2}\int \, dz_1 \, z_1 \int \,dz_2 \, z_2 \,M_{2,h_1 h_2 /P}^{i}(x,z_1,z_2,Q^2)=
(1-x)F^i_P(x,Q^2)\,. 
\end{eqnarray}
The dihadron fracture function  $M_{h_1,h_2/P}$ would be suitable, for example, 
for styding the coupled strange baryons/mesons production in the target fragmentation 
region, \textsl{i.e.} $h_1=\Lambda$ and $h_2=K$. It should be noted however that fragmentation 
functions in the strange channel are poorly known at present for one-hadron
fragmentation and unknown in the dihadron case.   

We wish conclude this sections by working out the general case.
Consider a n-inclusive Deep Inelastic Scattering process
$l+P\rightarrow l'+h_1 + h_2+\hdots h_n +X$ where the $n$ detected hadrons,
$h_i$, have all  $t_i\ll Q^2$, where $t_{i=1,\hdots n}=-(P-h_i)^2$ and 
fractional energy $z_i$ of the incoming hadron $P$. 
In analogy with eq.~(\ref{sigma_T1}), we may write the corresponding 
cross sections as 
\begin{equation}
\label{sigma_Tn}
\sigma_T=\int  \frac{du}{u} M^j_{h_1,\hdots h_n/P}(u,z_1,\hdots,z_n,Q^2)\; \hat{\sigma}_j(x/u,Q^2)\,.
\end{equation}
Comparing evolution equations for $M_{h/P}$ and $M_{h_1,h_2/P}$, we obtain, by induction, 
the $M^j_{h_1,\hdots h_n/P}(x,z_1,\hdots,z_n,Q^2)$ evolution equation. 
If  $M_{n_1/P}^j$ denotes the $n_1$-hadron fracture functions and $D^{n_2}_l$ the $n_2$-hadron 
fragmentation function, such that $n_1+n_2=n$, we get
\begin{eqnarray}
\label{Mn_evo}
&&Q^2\frac{\partial M^j_{n/P}}{\partial Q^2}(x,z_1,\hdots,z_n,Q^2)=\frac{\alpha_s(Q^2)}{2\pi}
\int_{\frac{x}{1-\sum_{k=1}^n z_k}}^{1}
\frac{du}{u}P_i^j(u)M^i_{n/P}(x/u,z_1,..,z_n,Q^2)+\frac{\alpha_s(Q^2)}{2\pi}
\sum_{q=1}^{n-1}\mathcal{P}_n  \{h,z\}\cdot\nonumber\\
&&\cdot\int_{\frac{x}{1-\sum_{k=q+1}^{n}z_k}}^{\frac{x}{x+\sum_{k=1}^{q}z_k}}
\frac{du}{u}\Big(\frac{u}{x(1-u)}\Big)^{q}\hat{P}_i^{jl}(u)M_{n-q}^i(x/u,z_{q+1},..,z_n,Q^2)
D_l^q \Big( \frac{z_1 u}{x(1-u)},.., \frac{z_q u}{x(1-u)},Q^2\Big)+\nonumber\\
&&+\frac{\alpha_s(Q^2)}{2\pi}\int_{x}^{\frac{x}{x+\sum_k^n z_k}}
\frac{du}{u}\Big(\frac{u}{x(1-u)}\Big)^n\hat{P}_i^{jl}(u)F^i_P(x/u,Q^2) 
D^n_l\Big( \frac{z_1 u}{x(1-u)},.., \frac{z_n u}{x(1-u)},Q^2\Big)\,.
\end{eqnarray}
The inhomogeneous term in the second line of eq.~(\ref{Mn_evo}) contains 
a permutation $\mathcal{P}_n\{h,z\}$ over hadrons indeces since we have assumed to measure 
$n$ distinct hadrons. The inner sum takes into account 
all the possible combinations $M_{n-q}\otimes D_q$ which give  
the $n$-hadrons final state configurations.  
It can be checked that the master formula, eq.~(\ref{Mn_evo}),  reproduces 
correctly the evolution equation for  $M_{h_1,h_2, h_3/P}$ when the latter 
is explicitely calculated as done for $M_{h_1,h_2/P}$ in the first part of this Section. 
Fair to say, this is a rather accademic exercise.  
Althought the ladder-type kinematics of LL Jet Calculus allows one to write a closed-form
evolution equation for $M_{n/P}$, the appeareance of an increasing number 
of unknown distributions prevents any further analysis.  

\section{Jet approach to initial state radiation}
\label{sec:jet_ISR}

The high multiplicity problem can of course be tackled by taking advantage 
of the some of the properties of QCD radiation. As is well known, hadron activity in a given 
hard interaction is often collimated in a definited portion of momentum space, 
this being a signature of the dominant collinear branching of pQCD dynamics.
For this reason jet cross-sections are the natural and, possibly, the most effective 
representation of hadronic final state. This approach avoids the introduction 
of multi-hadron distributions, which actually has caused the abandon of eq.~(\ref{Mn_evo}).   
Perturbative calculations with an arbitrary number of partons in the final state and 
experimental jet observables can be quantitatively compared 
only once a common jet-algorithm is chosen and used on both the theoretical and experimental level.
Let us sketch the jet approach to initial state radiation described 
in Ref.~\cite{kt_DIS}, in which the  $k_{\perp}$ clustering algorithm is used.   
The inclusive DIS structure functions $F_2$ can be decomposed in terms of $n$-particles 
exclusive cross-sections $F_2^{(n)}$ as \cite{detar}
\begin{equation}
\label{F2}
F_2(x,Q^2)=\sum_{n=1}^{\infty} F_2^{(n)}(x,Q^2)\,.
\end{equation}
If one moves from the exclusive $n$-particles cross-sections, eq.~(\ref{F2}), to the
 exclusive $n$-jets cross-sections, a factorized structure  emerges~\cite{kt_DIS}
\begin{equation}
\label{F2R}
F_2^{(n)}(x,Q^2;E_t^2,y_{\mbox{\small{cut}}})=\sum_{i=q,\bar{q}}\int_x^1 \frac{dz}{z}F_P^i(x/z,\mu_F^2)
R_{2,i}^{(n)}\Big(z,\alpha_s,\frac{Q^2}{E_t^2},y_{\mbox{\small{cut}}}\Big)\,,
\end{equation} 
where $y_{\mbox{\small{cut}}}$ represents the jet resolution parameter and 
defined in terms of an arbitrary perturbative scale $E_t^2$, with $\Lambda^2 \ll E_t^2 \leq Q^2$.
In eq.~(\ref{F2R}), initial state collinear divergences are absorbed into parton distributions 
functions, $F_P^i$. The jet-coefficients $R_{2,i}^{(n)}$ are calculable in perturbation theory
and, again, are jet-algorithm dependent. 
The $n$-jet cross-sections can be calculated by means of fixed order 
calculations. The main limitation of such an approach is represented by the 
technical difficulties of adding more and more partons in the final state. 
On the contrary, as shown in Ref.~\cite{kt_DIS}, a leading logarithmic accurate, 
Jet Calculus inspired, formulation of $n$-jet cross-sections is possible. 
In such an approach, initial state jets, in arbitrary numbers, are accounted for 
by using a generating functional method~\cite{kt_DIS}.
The $n$-jets cross-sections are then constructed by iterating the block-structure $G$,
\begin{eqnarray}
\label{block}
G_i^k(u,Q_i^2,Q_j^2)\equiv 
\Delta_i^j(Q_i^2,Q_j^2)\,\hat{P}_{j}^{lm}(u)\,J_m(Q_j^2,Q_0^2)\, \Delta_l^k(Q_l^2,Q_k^2)
\end{eqnarray}
\begin{figure}[ht]
\label{fig_block}
\begin{center}
\epsfig{file=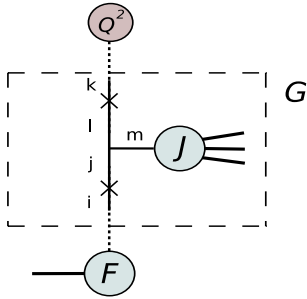,width=4cm,height=4cm,angle=0}
\caption{\small{Representation of one jet-insertion, eq.~(\ref{block}). The crosses represent 
Sudakov form factors, eq.~(\ref{Sudakov}). The three parton lines vertex indicates the 
real branching controlled by $\hat{P}(u)$. }}
\end{center}
\end{figure}\\
for each jet insertion and schematically represented in Fig.~3.
A Sudakov form factor $\Delta(Q_i^2,Q_j^2)$ \cite{Sudakov_FF}, defined as 
\begin{equation}
\label{Sudakov}
\Delta_i^j(Q_i^2,Q_j^2)\equiv \exp \Big[ - \sum_k \int_{Q_i^2}^{Q_j^2}
\frac{dt}{t}\int^{1-\frac{Q_i^2}{Q_j^2}}_{\frac{Q_i^2}{Q_j^2}} 
dz \frac{\alpha_s(t)}{2\pi} \hat{P}_i^{jk}(z) \Big]\,,
\end{equation} 
inhibites emissions off the
struck parton lines between the two scales $Q_i^2$ and $Q_j^2$ and thus 
guarantees that each jet is separated by the nearby jets by a inhibited-emission phase space region.
Hard parton emission in eq.~(\ref{block}) is then controlled by real splitting functions $\hat{P}(u)$ 
and its subsequent time-like evolution is taken into account via the jet function  $J(Q^2,k^2)$,
defined in Ref.~\cite{CT_res}, as
\begin{equation}
\label{jet_function}
J(Q^2,k^2)= \int_0^1 dz\, d(z,Q^2,k^2)\,.
\end{equation}
In eq.~(\ref{jet_function}), the distribution $d(z,Q^2,k^2)$ expresses the 
probability that an initial parton,
with mass $Q^2$, decays into a parton with a longitudinal momentum fraction
$z$ with respect to the parent parton with virtual mass $k^2 \ll Q^2$.
If  the intermediate partons mass $k^2$ is integrated over we 
get the parton fragmentation function $D$,
\begin{equation}
\label{Dfromd}
\int^{Q^2} dk^2\; d(z,Q^2,k^2)\equiv D(z,Q^2)\,.
\end{equation}
In eq.~(\ref{Dfromd}), both $d$ and $D$ would have only partonic indices 
since we are dealing with partonic jets distributions.
Such a formulation shares with the Jet Calculus approach 
the iterative construction. The $n$-jet cross-sections
are built using an alternate \textsl{allowed} and 
\textsl{prohibited parton emissions} pattern along the struck parton ordered virtualities, 
$Q_0^2< \hdots <Q_i^2<Q_j^2<Q_k^2<\hdots <Q^2$.
The main difference from the Jet Calculus approach resides 
in the exclusive formulation of eq.~(\ref{block}). 
The evolution function  $E_i^k(u,Q_i^2,Q_k^2)$, in eq.~(\ref{E_evo}), can be regarded however
as the analogous, at the inclusive level, of $G_i^k(u,Q_i^2,Q_k^2)$ in eq.~(\ref{block}). 
The former inclusively resums all parton emissions between the corresponding 
scales while the latter, instead, constrains such emissions to be approximately 
collimated in phase space.
\newpage
\section{Jet-like fracture functions}
\label{sec:Mjet}

The formalism of Ref.~\cite{kt_DIS} indeed can describe jets originating
from hard partons decays, whose emission off the active parton can be 
controlled at the perturbative level. 
The beam jet is excluded from such a decription. 
Its origin is mainly of soft and kinematical nature since
it results from the fragmentation of the spectator partons of the hadron remnants plus 
,eventually, semi-hard radiation coming from the evolution of the active parton. 
In the  $k_{\perp}$-algorithm the beam jet is therefore pre-clustered  
and not accounted for in the $n$-jet cross-sections. 
In view of the importance at present at forthcoming hadron colliders of 
describing such part of the process,   
we propose a new semi-inclusive jet-like distribution, $\mathcal{M}^i_{\sphericalangle}(x,Q^2,z,t)$, 
referring to it as to a jet-like fracture function. 
$\mathcal{M}^i_{\sphericalangle}(x,Q^2,z,t)$ 
expresses the probability of finding a parton $i$ with  
fractional momentum $x$ of the incoming hadron and virtuality $Q^2$, while 
a cluster of hadrons $h_i$ is detected in a portion of phase space $\mathcal{R}$
defined by two variables $z$ and $t$.
The phase space region $\mathcal{R}$ is limited by the constraint    
\begin{equation}
\label{t_constr}
\mathcal{R}: \;\; t_i = -(P-h_i)^2 < t,  \;\;\;\;    t_0 \le t \le Q^2\;.
\end{equation}
Once the clustering procudere is performed, the variable $z$ is obtained 
by summing the fractional longitudinal momenta
of all hadrons $h_i$ satisfying the phase space constraint, eq.~(\ref{t_constr}):
\begin{equation}
\label{z_def}
z=\sum_i z_i, \;\;\; h_i \in \mathcal{R}\,.  
\end{equation}
In analogy with standard inclusive DIS, which makes use of parton distributions 
functions, we may write
\begin{equation}
\label{Mjet_def=sigma}
\frac{1}{\sigma_{tot}}\frac{d\sigma}{dx dQ^2 dz dt}\propto x \sum_{i=q,\bar{q}}~ e_i^2 ~
\mathcal{M}^i_{\sphericalangle}(x,Q^2,z,t)\,.
\end{equation}
As in the inclusive case, $x$ and $Q^2$ are fixed by 
the scattered lepton kinematics. 
By defining the $n$-particle exclusive cross-sections, 
which may be obtained directly from experiments, as
\begin{equation}
\Sigma_{excl}^{(n)}\equiv \frac{1}{n!} \frac{d^{2n+2}\sigma^{(n)}}{dx dQ^2 \prod_{m=1}^{n}dz_m dt_{m}}
\end{equation}
we may construct the distributions in eq.~(\ref{Mjet_def=sigma}) by implementing 
the phase space constraints, eq.~(\ref{t_constr}):
\begin{equation}
\label{Mjet_def}
\frac{1}{\sigma_{tot}}\frac{d\sigma}{dx dQ^2 dz dt}\equiv
\frac{1}{\sigma_{tot}}
\sum_{k=1}^{\infty} \Big\{ \prod_{m=1}^{k} \int^{t}_{t_0} dt_{m} 
\int_0^1  dz_m \Big\} \, \Sigma_{excl}^{(k)}
\, \delta\Big({z-\sum_{k=1}^{n}z_k}\Big)\,,
\end{equation}
where $t_0$ corresponds  to beam pipe acceptance loss.
At the dynamical level, the active parton described by $\mathcal{M}_{\sphericalangle}^i$
increases its virtuality toward the hard vertex in a series of 
subsequent branchings. Futhermore, at leading logarithmic accuracy, 
each emission is strongly ordered, with the softest $k_{\perp}$-emissions close to the proton remnant.
The invariant momentum transfer constraint in eq.~(\ref{t_constr}) thus perfectly matches this leading
logarithmic picture, including soft and semi-hard radiation into $\mathcal{M}_{\sphericalangle}^i$ 
up to the scale $t$. All log-enhanced emissions above that scale 
are then resummed by using the function $E$, eq.~(\ref{E_evo}). 
Moreover, at variance with the full inclusive case, the parton initiating 
the space-like cascade can be specified by inspecting initial state radiation itself. 
It has a fractional momentum $1-z$, where $z$ is overall fractional momentum taken 
away by the hadrons with $t_i \leq t$ and the highest allowed virtuality, $t$, 
according to strong ordering. 
When $t$ is chosen in the perturbative region we may 
write, according to Jet Calculus,
\begin{equation} 
\label{Mdef}
\mathcal{M}_{\sphericalangle}^j(x,Q^2,z,t)=\int_x^{1-z} \frac{dw}{w} 
\mathcal{M}_{\sphericalangle}^i(w,t,z,t) ~ E_i^j(x/w,t,Q^2)\,.
\end{equation}
\begin{figure}[ht]
\label{fig:Mjet}
\begin{center}
\epsfig{file=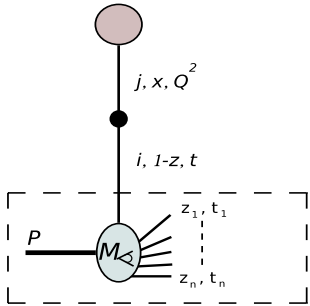,width=4cm,height=4cm,angle=0}
\caption{\small{Representation of the jet-like fracture function $\mathcal{M}_{\sphericalangle}^i$
as in eq.~(\ref{Mdef}).
The black blob represents the inclusive parton emissions between virtuality $t$ and $Q^2$, 
the bright blob on top the hard scattering process.
Also indicated are the space-like parton momentum fractions. The forward radiation off 
 $\mathcal{M}_{\sphericalangle}^i$,
satisfying the constraint eq.~(\ref{t_constr}), is shown.}}
\end{center}
\end{figure}
The right hand side of eq.~(\ref{Mdef}) is depicted in Fig.~4.
Convolution limits are fixed by requiring momentum conservation.
By differentiating eq.~(\ref{Mdef}) with respect to $Q^2$,
we obtain the evolution equations for $\mathcal{M}_{\sphericalangle}^j$ which reads
\begin{equation}
\label{Mjet_evo}
Q^2\frac{\partial}{\partial Q^2}\mathcal{M}_{\sphericalangle}^j(x,Q^2,z,t)=\frac{\alpha_s(Q^2)}{2\pi}
\int_{\frac{x}{1-z}}^1\frac{du}{u}P_k^j(u)\mathcal{M}_{\sphericalangle}^k(x/u,Q^2,z,t)\,.
\end{equation}
This equation describes how the virtual photon resolves 
the distributions  $\mathcal{M}_{\sphericalangle}^j$, 
when the virtuality of the latter is varied. 
As already stated, the evolution equation, eq.~(\ref{Mjet_evo}), 
actually resums large logarithm of the type  $\alpha_s^n \log^n (Q^2/t)$.
In real processes, $t$-ordering is only partially realized. Higher order 
corrections produce partons that, even if originated by a parent parton 
with a hard $\widetilde{t} \geq t$ scale, 
could give, at the end of the time-like shower, final state hadrons with $t_i \le t$.
The clustering procedure of course does not distinguish 
the origin of such hadrons, nor the virtuality of the parent parton emitted 
off the space-like chain. 
All of them are however included in  $\mathcal{M}_{\sphericalangle}^i$, 
according to eq.~(\ref{t_constr}).
The description of  $\mathcal{M}_{\sphericalangle}^j$ therefore becomes increasingly 
reliable as much as the accuracy in space-like partonic shower is enhanced. 
This, of course, can be achieved by inserting appropriate higher loop splitting functions
\begin{equation}
P_k^j(u)=P_k^{j\,(0)}(u)+\frac{\alpha_s}{2\pi}P_k^{j \, (1)}(u)+\hdots
\end{equation}
in the evolution equations, eq.~(\ref{Mjet_evo}). 
We also note that the latter is formally equivalent to the one for 
one-particle inclusive extended fracture functions of Ref.~\cite{camici}. 
This formal equivalence is expected since we may consider 
the hadrons contained in  $\mathcal{R}$ as a pseudo-particle specified by 
fractional longitudinal momentum $z$ and invariant momentum transfer $t$.
At the experimental level, the proposed distributions 
have been already adopted in diffractive DIS measuraments at HERA, see Refs.
\cite{Mjet&H1,Mjet&ZEUS}. When the diffractive event is tagged by observing a rapidity gap, 
the unmeasured low-mass proton excitation
and eventually soft $p_{\perp}$ fragments are taken into account by using an integrated 
distribution as $\mathcal{M}_{\sphericalangle}^j$. 
Of particular interest is the issue concerning the  $\mathcal{M}_{\sphericalangle}^j$
factorization properties and 
we will discuss it in the remaining part of this Section. 
Factorization of ordinary fracture functions were demonstrated in DIS in 
the single-particle case in Refs.~\cite{Fact_M_coll,Fact_M_soft}. 
Since $\mathcal{M}_{\sphericalangle}^j$ is actually more inclusive than 
ordinary extended fracture functions \cite{camici},
for which factorization holds, we does not expect any factorization breaking effect in DIS.
As a result, as for standard parton distribution in inclusive processes,
factorization guarantees that once the beam jet and the forward radiation 
are measured in a given experiment and assigned to $\mathcal{M}_{\sphericalangle}^j$ , 
such a distribution can be used in a different experiment solving $\mathcal{M}_{\sphericalangle}^j$ 
evolution equations.
Morover factorization allows a generalization of eq.~(\ref{F2R}) to include 
also the beam jet in the $n$-jet cross-sections, 
by simply substituting a parton distribution with a jet-like fracture function: 
\begin{equation}
\label{F2R_Mjet}
F_2^{(n+1)}(x,Q^2;E_t^2,y_{\mbox{\small{cut}}})=\sum_{i=q,\bar{q}}\int_0^1 dz 
\int_{\frac{x}{1-z}}^1 \frac{du}{u}
\mathcal{M}_ {\sphericalangle}^i(x/u,Q^2,z,y_{\mbox{\small{cut}}} E_t^2)\,
R_{2,i}^{(n)}\Big(u,\alpha_s,\frac{Q^2}{E_t^2},y_{\mbox{\small{cut}}}\Big)\,.
\end{equation}
In eq.~(\ref{F2R_Mjet}), the fraction $z$ of the beam jet, being not measured, is integrated over. 
The scale $t$ is set to $t \simeq y_{\mbox{\small{cut}}} E_t^2$ and $\mathcal{M}_{\sphericalangle}^i$ 
should be properly evolved 
according to eq.~(\ref{Mjet_evo}) before being inserted in eq.~(\ref{F2R_Mjet}).
Turning now to hadron-hadron collisions, consider a semi-inclusive Drell-Yan type process: 
\begin{equation}
p+p \rightarrow C_1 + C_2 + \gamma^*_{~\hookrightarrow~ l^+ l^-} + X\,. 
\end{equation}
The invariant mass of the lepton-pair provides 
the perturbative hard scale and the leptonic final state
allows a clean inspection of QCD initial state radiation. 
$C_1$ and $C_2$ are two hadronic forward clusters,  
containing  each a beam remnant jet and associated radiation,  
and are defined in the phase space regions  $\mathcal{R}_1$ and  
$\mathcal{R}_2$, eq.~(\ref{t_constr}), limited by $t_1$ and $t_2$\,.
If we assume for a while that factorization holds, the cross-sections for the process can be written as 
\begin{equation}
\label{DY_Mjet}
\frac{d\sigma^{DY}}{dt_1 dt_2 dQ^2 dz_1 dz_2}=\sum_{i,j=q,\bar{q}}\int\int dx_1 dx_2 
\Big( \mathcal{M}_{\sphericalangle}^{i}(x_1,Q^2,z_1,t_1)
\mathcal{M}_{\sphericalangle}^{j}(x_2,Q^2,z_2,t_2) + i\leftrightarrow j   \Big)~\delta(s-x_1 x_2 Q^2)\,,
\end{equation}
in complete analogy with the inclusive Drell-Yan case, for which instead factorization 
has been proven in Refs.~\cite{DY_factorization_1,DY_factorization_2}.
Eq.~(\ref{DY_Mjet}), pictorially represented in Fig.~5, does not take into 
account possible multiple hard interaction, 
modelled recently in Ref.~\cite{Sjostrand}, nor any model for the soft 
remnant-remnant interaction, which can be instead modelled as in Ref.~\cite{DPM}. 
Eq.~(\ref{DY_Mjet}) can be used to measure the strenght of factorization 
breaking effects in semi-inclusive hadron-hadron collisions 
since factorization properties of the cross sections 
could be connected with the observed forward radiation pattern.
\begin{figure}[h]
\label{fg4}
\begin{center}
\epsfig{file=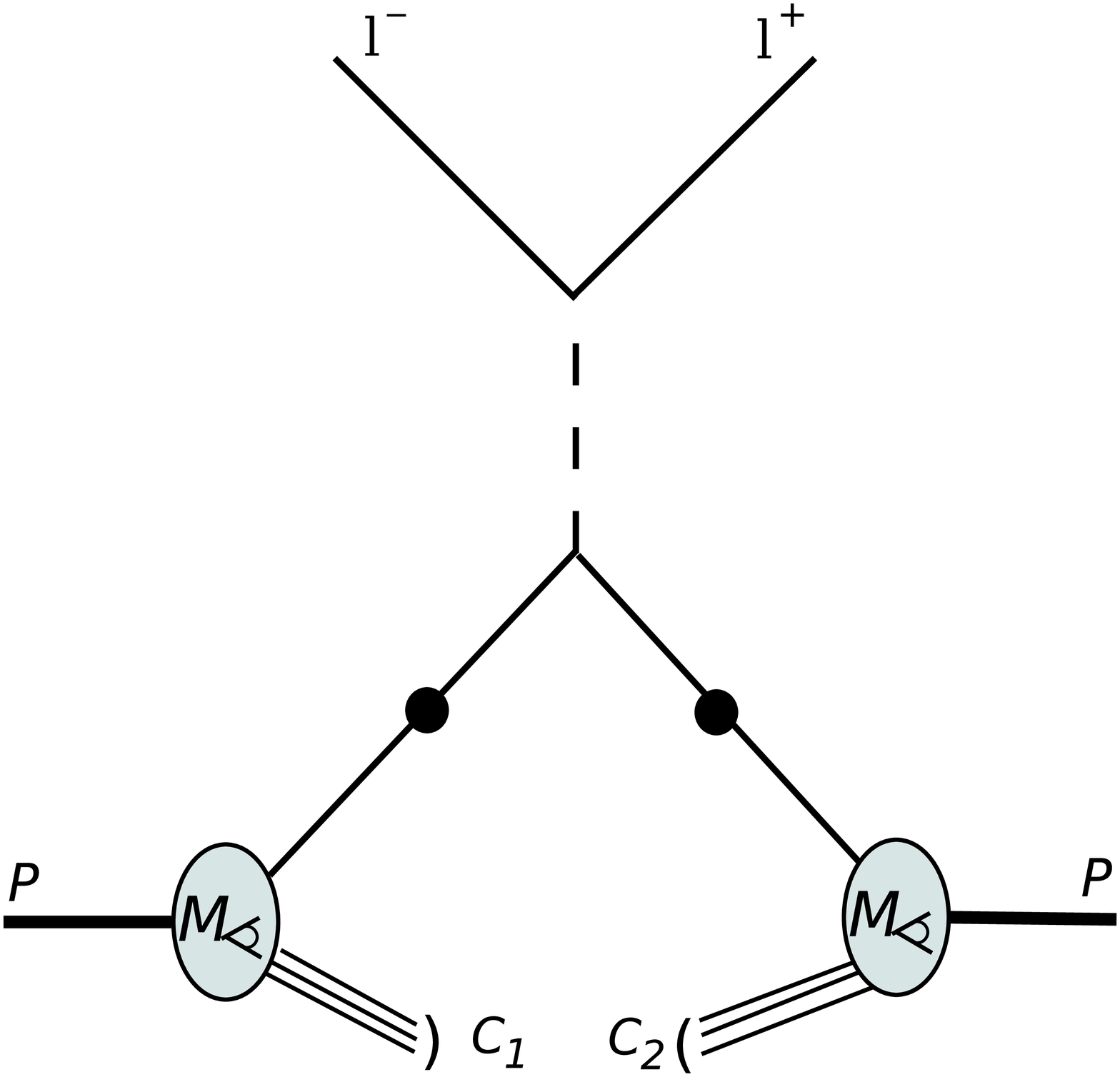,width=7cm,height=6cm,angle=0}
\caption{\small{Representation of a semi-inclusive Drell-Yan process, as in eq.~(\ref{DY_Mjet}).
The dashed line indicates the virtual boson.}}
\end{center}
\end{figure}

\section{Conclusions}
\label{sec:end}

In the present work we have developed an alternative 
QCD-based approach to initial state radiation in hard, 
hadron-initiated process by using
the Jet Calculus and Fracture Function formalism. 
We propose novel jet-like Fracture Functions which 
depend explicitely on a minimal set of variables describing forward 
radiation and target remnants. 
The lack of knowledge on soft hadronic dynamics 
compels the use of non-perturbative distributions.
As in the inclusive case, 
the $Q^2$-evolution of the proposed semi-inclusive multi-particle distributions
can be predicted in QCD. As a result, an explicit description of the QCD initial state radiation becomes possible. 
In this paper we limit ourselves to results within leading logarithmic approximation. 
We do not foresee however any serious limitation in implementing 
higher order corrections and eventually coherence effects in the formalism presented here.  
Let us add that, since the issue of QCD factorization in hard 
hadron-hadron reactions is closely connected with the pattern of soft and semi-hard initial 
state radiation, this novel approach may result in a useful framework to reconsider it.
This subject is at the moment under scrutiny as well as further possible applications of the formalism to the dynamics of the minimum bias and of the underlying event at hadron colliders. 

\begin{center}

\end{center}
\end{document}